\documentclass[1p,times]{elsarticle}
\usepackage{graphicx}
\usepackage{color}
\usepackage{amssymb}
\usepackage{amsmath}
\usepackage{xspace}
\usepackage{multirow}   
\usepackage{booktabs}   
\usepackage{longtable}
\usepackage{xcolor}

\usepackage{hyperref}
\hypersetup{colorlinks=true, linkcolor=black, citecolor=black, urlcolor=black}
\usepackage{lineno}

\biboptions{comma, square, sort&compress}
\journal{Communications in Nonlinear Science and Numerical Simulation} 
\begin{document}
\begin{frontmatter}



\ead{thomas.schrefl@donau-uni.ac.at}
\cortext[cor1]{Corresponding author} 

\title{Exploring the hysteresis properties of nanocrystalline permanent magnets using deep learning}
\author[a,b]{Alexander Kovacs}
\author[c,d,e]{Lukas Exl}
\author[a,b]{Alexander Kornell}
\author[a,b]{Johann Fischbacher}
\author[a,b]{Markus Hovorka}
\author[a,b]{Markus Gusenbauer}
\author[b]{Leoni Breth}
\author[b]{Harald Oezelt}
\author[f]{Masao Yano}
\author[f]{Noritsugu Sakuma}
\author[f]{Akihito Kinoshita}
\author[f]{Tetsuya Shoji}
\author[f]{Akira Kato}
\author[a,b]{Thomas Schrefl\corref{cor1}}

\address[a]{Christian Doppler Laboratory for Magnet design through physics informed machine learning, Viktor Kaplan-Stra{\ss}e 2E, 2700 Wiener Neustadt, Austria}
\address[b]{Department for Integrated Sensor Systems, Danube University Krems, Viktor Kaplan-Stra{\ss}e 2E, 2700 Wiener Neustadt, Austria}
\address[c]{Department of Mathematics, University of Vienna, Oskar-Morgenstern-Platz 1, 1090 Vienna, Austria}
\address[d]{Wolfgang Pauli Institute, Oskar-Morgenstern-Platz 1, 1090 Vienna, Austria}
\address[e]{Research Platform MMM Mathematics-Magnetism-Materials, Oskar-Morgenstern-Platz 1, 1090 Vienna, Austria}
\address[f]{Advanced Materials Engineering Div., Toyota Motor Corporation, 1200, Mishuku Susono, Shizuoka 410-1193 Japan}


\begin{abstract}
We demonstrate the use of model order reduction and neural networks for estimating the hysteresis properties of nanocrystalline permanent magnets from microstructure. With a data-driven approach, we learn the demagnetization curve from data-sets created by grain growth and micromagnetic simulations. We show that the granular structure of a magnet can be encoded within a low-dimensional latent space. Latent codes are constructed using a variational autoencoder. The mapping of structure code to hysteresis properties is a multi-target regression problem. We apply deep neural network and use parameter sharing, in order to predict anchor points along the demagnetization curves from the magnet's structure code. The method is applied to study the magnetic properties of nanocrystalline permanent magnets. We show how new grain structures can be generated by interpolation between two points in the latent space and how the magnetic properties of the resulting magnets can be predicted. 
\end{abstract}

\end{frontmatter}


\section{\label{sec_introduction}Introduction}
The macroscopic properties of permanent magnets arise from the interplay of intrinsic material properties and the magnet's granular microstructure. Traditionally, structural characterization \cite{grossinger2005structural}, imaging and magnetic measurements \cite{manaf1993microstructure,bernardi2000preparation} as well as micromagnetic theory \cite{kronmuller1987theory} and  simulations \cite{schrefl1994remanence,griffiths1998computer,tsukahara2020relationship} have been applied to obtain a better understanding of the impact of microstructure on the magnetic properties. The relation between structure and properties can be treated as mapping between an input space and an ouput space. In this work, we apply a data-driven approach to map the microstructure of a nano-crystalline magnet to its demagnetization curve. The data is derived from a sequence of solutions of  nonlinear partial differential equations. Alternatively, experimental data such as structural images and measured demagnetization curve could be used to train the machine learning model.

Key properties of permanent magnets are the remanent magnetization $M_\mathrm{r}$, which is the magnetization at zero field $M_\mathrm{r} = M(H_\mathrm{ext}=0)$, and the coercive field $H_\mathrm{c}$, which is the field that gives zero magnetization $M(H_\mathrm{ext}=H_\mathrm{c}) = 0$. 
In a permanent magnets the grains are not perfectly aligned. The misorientation angle of a grain denotes the angle between the grain's anisotropy axis and the alignment direction. In nanocrystalline magnets, the coercive field decreases with increasing standard deviation of the misorientation angle \cite{kronmuller1999magnetization}. Similarly, remanence decreases with increasing misorientation. Both remanence and coercivity depend on the grain size. When the exchange coupling between the grains is zero or weak, the coercivity of a magnet decreases with increasing grain size. This is attributed to non-uniform demagnetizing fields near the edges of the polyhedral grains  \cite{gronefeld1989calculation,bance2014grain}. In isotropic, nano-crystalline magnets with strong ferromagnetic exchange coupling between the grains the situation is different: Coercivity decreases with decreasing grain size \cite{manaf1991enhanced,fukunaga1992effect}. Exchange interactions dominate over the local magneto-crystalline anisotropy \cite{callen1977initial,herzer1990grain} causing partial domain walls at grain boundaries reduce the switching field \cite{schrefl1994remanence}. At the same time, these partial domain walls enhance the remanence, and the remanent magnetization is found to increase with decreasing grains size. 

Traditionally, micromagnetic theory are used for forward simulation. For a few given grain structures the demagnetization curve, the remanence, and the coercivity are computed. In this paper we introduce a different approach. We use micromagnetic simulations to train a machine learning model. Once such a model is trained and ready to use, hysteresis properties can be estimated on the fly without the need of time-consuming simulation. Such a tool opens new possibilities for magnet design. On obvious application is to explore a wide range of different magnetic microstructures, because it literally takes no time to compute the hysteresis properties. However, machine learning offers even more possibilities. As part of our machine learning tool chain we use a generative model. Generative models can generate new samples within the input space which differ from the original training set. They can be used to create new samples that are different from training data. One prominent example is the use of variational autoencoders for face morphing  \cite{yeh2016semantic,foster2019generative}. Autoencoders learn to copy
their inputs to their outputs.  Thereby they learn representing the input
state in lower dimensionality: An encoder model decodes the input image to a low dimensional latent code, and a decoder models reconstructs the original image. In face morphing, new images of faces are generated by linear interpolation between two points in the latent space. Similar ideas can be applied in computational magnetics. From points in the latent space, new microstructures of permanent magnets can be generated and the associated hysteresis properties can be predicted. 

We applied a convolutional autoencoder to reduce the dimension of the input space and a multi-headed  neural network regressor to learn a relation between latent space and magnetic properties. The work presented in this paper is a proof of concept that such a mapping between structure and properties can be learned using neural networks.

Previously, Lu and co-workers solved parametric partial differential equations using DeepONet \cite{lu2019deeponet}, a neural network with special architecture for the approximation of non-linear operators. Their method is based on the universal operator approximation theorem \cite{chen1995universal}, which states that a neural network with a single hidden layer can accurately approximate nonlinear continuous operators that map between compact subsets of certain function spaces. The nonlinear operator $G$, the mapping, depends on a general function $u(x)$ and transforms any input $y$ into an output $G(u)(y)$. Once trained, the network approximates the mapping for previously unseen functions $u(x)$. Lu and co-workers \cite{lu2019deeponet} use a discrete representation of the function $u$, which can be the space dependent coefficient of a partial differential equation.  Bhattacharya and co-workers \cite{Bhattacharya2021} combined model reduction and neural networks for mapping between infinite input and output spaces. In particular, they apply their method to map between the coefficient and the solution of an elliptic partial differential equation. The method is based on the principal component analysis dimension reduction of the input and output spaces. Then a neural network is applied to map between the dimension-reduced spaces. Kovacs and co-workers \cite{kovacs2019learning} applied an convolutional autoencoder to reduce the dimension of the solution of the Landau-Lifshitz-Gilbert equation, which describes the time evolution of the magnetization. Then they use a data-driven approach to learn the time evolution in the reduced dimensions. With this approach they can predict the magnetization dynamics of thin film elements for previously unseen external fields. As an alternative to neural networks, kernel methods have been used to learn the solution of the Landau-Lifshitz-Gilbert equation in latent space \cite{exl2020learning,exl2021prediction}.  

\begin{figure}[!htb]
	\centering
	\includegraphics[scale=0.5]{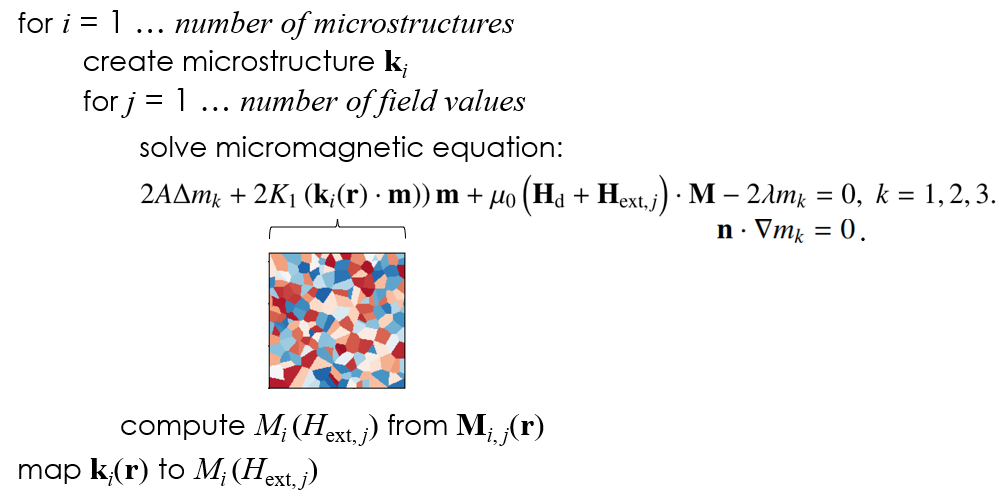}
	\caption{\label{fig_learning}Learning demagnetization curves from microstructure. Within the supervised training process, for each microstructure $i$ a sequence of solutions $j$ is mapped to a discrete representation of the demagnetization curve. For Brown's micromagnetic equation please see section \ref{sec_brown}.} 
\end{figure}

In this work we present a neural network based methodology to predict the demagnetization curve of nanocrystalline permanent magnets from the microstructure. Nanocrystalline permanent magnets have been the subject of intensive research both experimentally \cite{hadjipanayis1988magnetic,manaf1991enhanced,manaf1993microstructure} and numerically \cite{fukunaga1992effect,schrefl1994remanence,rave1997micromagnetic,griffiths1998computer}. Here we use micromagnetic simulations to calculate the magnetization curve of microstructures generated by Voronoi-tesselation. The demagnetization curve gives the total magnetization projected onto the direction of the external field 
\begin{equation}
	M(H_\mathrm{ext}) = \frac{1}{V} \int_V \mathbf{M(\mathbf{r})} \cdot \frac{\mathbf{H}_\mathrm{ext}}{H_\mathrm{ext}} dV,
\end{equation}
with $H_\mathrm{ext} = \left| \mathbf{H}_\mathrm{ext} \right|$. The integral is over the volume $V$ of the magnet. The magnetization $\mathbf{M}(\mathbf{r})$ may change with position $\mathbf{r}$. For a given external field $\mathbf{H}_\mathrm{ext}$, the magnetization distribution follows from the solution of Brown's micromagnetic equation. Magnetic materials show hysteresis. Thus, the solution is not unique. To compute the demagnetization curve, the equilibrium distribution of the magnetization for a sequence of decreasing external fields has to be found \cite{kinderlehrer1997hysteretic}. The microstructure and the external field are reflected by space-dependent and uniform coefficients of the partial differential equation, respectively. We apply a variational autoencoder to represent the microstructure in a low dimensional function space, represent the demagnetization  curve by a few discrete anchor points, and learn a map between the microstructure in reduced space and the anchor points. Figure \ref{fig_learning} introduces our data driven approach to learn the demagnetization curve from the microstructure of a magnet.

This work is a proof of concept for machine learning hysteresis from microstructure. We show how to build and train a machine learning model that can predict hysteresis properties from microstructure images. Further, we demonstrate how methods from generative deep learning may be applied for magnet development. The paper is organized as follows. In section \ref{sec_micromagnetic}, we summarize the micromagnetic background and show how we compute the demagnetization curve for a given microstructure. In section \ref{sec_ml}, we introduce the basic concepts of machine learning applied in our work such as dimensionality reduction using variational autoencoders and multi-target neural network regression. In section \ref{sec_results_hyst}, we present results for learning demagnetization curves of nano-crystalline permanent magnets. In section \ref{sec_results_latent}, we show examples of latent space interpolation applied to microstructures of nanocrystalline magnets.  In section \ref{sec_outlook}, we summarize our work and give a perspective for the future application of the presented computational tool chain. 

\section{\label{sec_micromagnetic}Micromagnetic background}

\subsection{\label{sec_brown}Brown's micromagnetic equations}

The stable or meta-stable magnetization distribution of a magnet is given by the solution Brown's micromagnetic equation \cite{brown1963micromagnetics} which follows from the variation of the Gibbs free energy. For permanent magnets the Gibbs free energy $E$ is the sum of the ferromagnetic exchange energy $E_\mathrm{ex}$, the magnetocystalline anisotropy energy $E_\mathrm{ani}$, the magnetostatic energy $E_\mathrm{mag}$, and the Zeeman energy $E_\mathrm{ext}$:
\begin{eqnarray}
	E_\mathrm{ex} & = & \int_V A(\mathbf{r}) \left[ 
	\left( \nabla m_1(\mathbf{r}) \right)^2 +
	\left( \nabla m_2(\mathbf{r}) \right)^2 +
	\left( \nabla m_3(\mathbf{r}) \right)^2 \right] dV, \\
	E_\mathrm{ani} & = & - \int_V K_1(\mathbf{r}) \left( \mathbf{k(\mathbf{r})} \cdot \mathbf{m}(\mathbf{r})\right)^2 dV,\\
	E_\mathrm{mag} & = & - \frac{\mu_0}{2} \int_V \mathbf{H}_\mathrm{d}(\mathbf{r}) \cdot \mathbf{M}(\mathbf{r}) \mathrm dV, \\
	E_\mathrm{ext} & = & - \mu_0\int_V \mathbf{H}_\mathrm{ext}(\mathbf{r}) \cdot \mathbf{M}(\mathbf{r}) \mathrm dV. 
\end{eqnarray}
Here $\mathbf{m}(\mathbf{r}) = \left(m_1, m_2, m_3 \right)^\mathrm{T} = \mathbf{M}/M_\mathrm{s}$ is the unit vector parallel to the magnetization $\mathbf{M}(\mathbf{r})$;  $M_\mathrm{s}(\mathbf{r})$ is the spontaneous magnetization; $A(\mathbf{r})$ is the exchange constant, $K_1(\mathbf{r})$ is the uniaxial anisotropy constant, $\mathbf{k}(\mathbf{r})$ is the unit vector along the anisotropy direction;  $\mu_0$ is the permeability of vacuum; $\mathbf{H}_\mathrm{d}(\mathbf{r})$ is the self-demagnetizing field; and $\mathbf{H}_\mathrm{ext}$ is the external field. The variation of the total energy with respect to $m_k$ \cite{kronmuller2003micromagnetism} gives the partial differential equation and boundary conditions for the equilibrium state. These are
\begin{equation}
\label{eq_brown}
2A\Delta m_k + 2 K_1 \left( \mathbf{k}(\mathbf{r}) \cdot \mathbf{m})\right) \mathbf{m} + \mu_0 \left( \mathbf{H}_\mathrm{d} + \mathbf{H}_\mathrm{ext}\right) \cdot \mathbf{M} - 2 \lambda m_k = 0, \; k = 1, 2, 3.
\end{equation}
in the volume, and 
\begin{equation}
	\label{eq_brown_bnd}
	\mathbf{n} \cdot \nabla m_k = 0
\end{equation}
on the surface with surface normal $\mathbf{n}$. The last term contains the Lagrange paramter $\lambda$ associated with the constraint $m_1^2 + m_2^2 + m_3^2 = 1$. This constraint ensures that only the direction, not the length, of the magnetization vector varies. A solution of (\ref{eq_brown}) and (\ref{eq_brown_bnd}) corresponds to a stationary point of the Gibbs free energy. For finding minima, we apply the curvilinear steepest descent method \cite{goldfarb2009curvilinear,exl2014labonte} as implemented in the micromagnetic solver fidimag \cite{bisotti2018fidimag,Fidimag}. 

\subsection{\label{sec_grains}Grain structure generation and computation of demagnetization curves}

In melt-spun nanocrystalline magnets the granular structure is reflected by the anisotropy direction $\mathbf{k}(\mathbf{r})$ which varies from grain to grain. All other material parameters are constant. 

Similar to previous micromagnetic studies \cite{schrefl1994remanence,rave1997micromagnetic} of nanocrystalline magnets we use two-dimensional structures. The easy axes of the grains as well as the magnetization are constrained to a plane. The magnetization is taken to be uniform in the direction perpendicular to this plane. For computing the equilibrium magnetic states $\mathbf{M}(\mathbf{r})$ we use the finite difference micromagnetic solver fidimag \cite{bisotti2018fidimag}. The user guide \cite{Fidimag} comes with an example to generate two-dimensional grain structures based on Voronoi-tesselation. We use the software of this example to generate the granular structures. For all structures we fix the size to $320 \times 320$~nm$^2$. We randomly sampled the number of grains $n_\mathrm{g}$ and the maximum misorientation angle $\theta$ uniformly from $\left[16,256\right]$ and $\left[-\pi,\pi\right]$, respectively. For each grain structure the $n_\mathrm{g}$ angles of the easy directions were taken uniformly from the interval $\left[-\theta,\theta\right]$. 

In the plane of the easy axes, the size of the computational cells was 2 nm. To mimic translational symmetry in the direction perpendicular to the plane containing all easy axes, the cell dimension was set to $2 \times 2 \times 36000$~nm$^3$.    

From the computed demagnetization curve $M(H_\mathrm{ext})$ we extract magnetization and field values that define the following six anchor points:
\begin{enumerate}
\item
The magnetization at the anisotropy field $M_1 = M(2 K_1/(\mu_0 M_s))$, 
\item
The remanent magnetization $M_\mathrm{r} = M(0)$, 
\item
The knee field $H_\mathrm{k,90}$ defined by $M(H_\mathrm{k,90}) = 0.9 M_\mathrm{r}$,
\item
$H_\mathrm{k,80}$ defined by $M(H_\mathrm{k,80}) = 0.8 M_\mathrm{r}$, and
\item
$H_\mathrm{k,70}$ defined by $M(H_\mathrm{k,70}) = 0.7 M_\mathrm{r}$.
\item
The coercive field $H_\mathrm{c}$ defined by $M(H_\mathrm{c}) = 0$,
\end{enumerate}
The remanent magnetization $M_\mathrm{r}$, the coercive field $H_\mathrm{c}$, and the knee field $H_\mathrm{k,90}$ are key figures of merit for a permanent magnet \cite{strnat1978rare}. The critical fields $H_\mathrm{k,80}$ and $H_\mathrm{k,70}$ were defined in analogy to the classical knee field $H_\mathrm{k,90}$. With the above defined six anchor points along the demagnetization curve it is possible to approximate the demagnetization curve of permanent magnets. 

\section{\label{sec_ml}Machine learning background}

\subsection{Dimensionality reduction with variational autoencoders}

Deep neural networks are currently widely used in object recognition, image creation, machine translation, and speech recognition \cite{russell2020artificial}. Neural networks show advantages as compared to classical machine learning algorithms for problems involving high dimensional data like images. Neural networks are trained by examples of input-ouput pairs $(\mathbf{x}_l,\mathbf{y}_l)$. Here $\mathbf{x}$ is the input vector, for example the $x$-components of the anisotropy direction on the computational grid; $\mathbf{y}$ is the output vector, for example the values that define the anchor points along the demagnetization curve $(M_1, M_\mathrm{r}, H_\mathrm{k,90}, H_\mathrm{k,80}, H_\mathrm{k,70}, H_\mathrm{c} )^\mathrm{T}$.  The index $l$ runs from 1 to the total number of training pairs $n$. The neural network approximates the output vector
\begin{equation}
  \mathbf{y}_\mathrm{approx}(\mathbf{x}) = \mathcal{N}\left( \mathbf{x}, \mathbf{w}\right).
\end{equation}   
The vector $\mathbf{w}$ represents the weights and biases of the network. The weights and biases are the learnable parameters of the network which are determined during training of the network by minimizing the mean squared error (MSE) loss function
\begin{equation}
	L = \frac{1}{n} \sum_{l=1}^n \left( \mathbf{y}_{\mathrm{approx},l} - \mathbf{y}_l  \right)^2.
\end{equation}

\begin{figure}[!tb]
	\centering
	\includegraphics[scale=0.33]{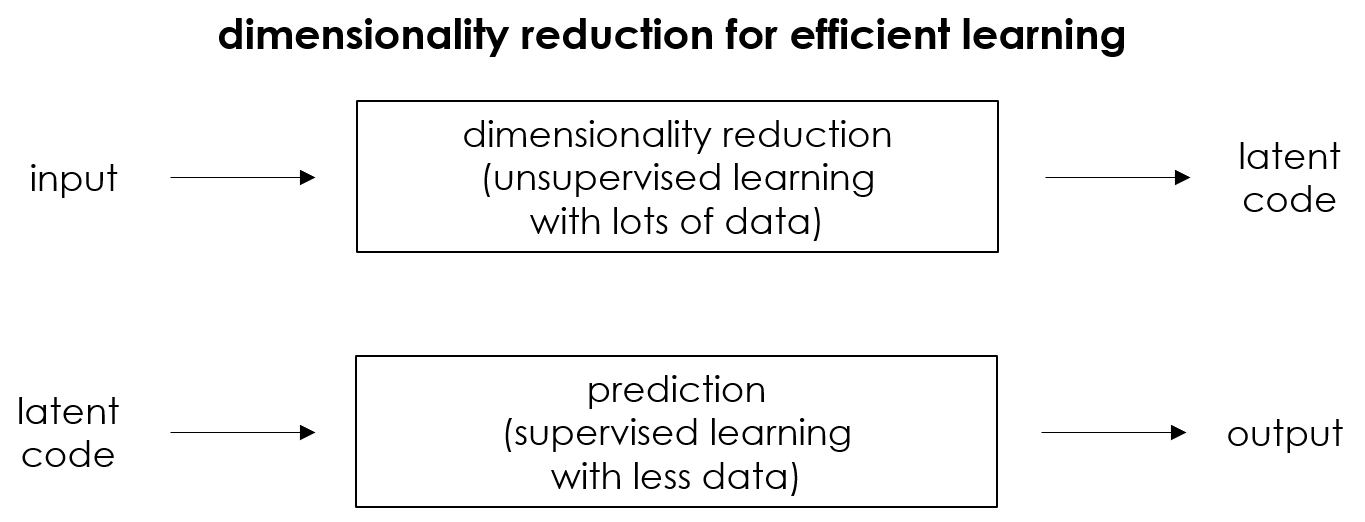}
	\caption{\label{fig_super}Using a low-dimensional latent code to automate feature selection in situations where the number of labeled data points is small (after \cite{buduma2017fundamentals}).} 
\end{figure}

\begin{figure}[!tb]
	\centering
	\includegraphics[scale=0.52]{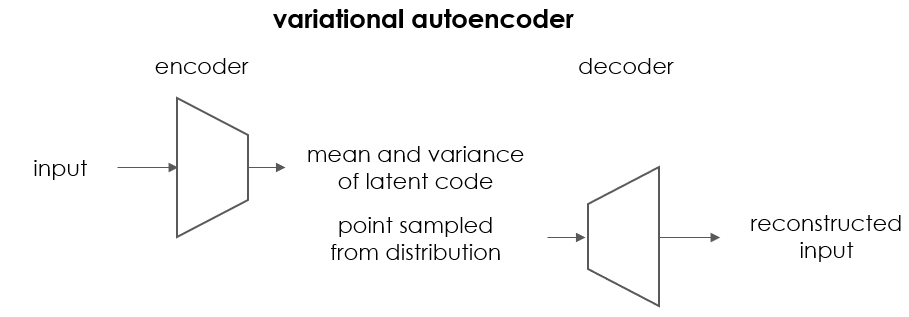}
	\caption{\label{fig_auto} A variational autoencoder consists of an encoder and a decoder. The encoder maps the high-dimensional input to a distribution in latent space represented by the mean and the variance of the latent code. The decoder reconstructs the input from a latent point randomly sampled from the distribution given by the mean and the variance.} 
\end{figure}

The number of trainable parameters increases with the dimensions of the input vector. Large models with many hidden layers can express nonlinear mappings and generalize to unseen data. This is considered to be the key advantage of deep neural networks. However, training of large networks requires a huge amount of data. In materials science the number of input-output pairs $n$ is small. In this work, the calculation of demagnetization curves, which is time consuming, is the bottleneck that limits the size of the training set. Generating microstructures is much cheaper than computing the demagnetization curve. We can afford to have a lot of unlabeled data, the grain structures sampled as described in section \ref{sec_micromagnetic}. On the other hand, we only have a limited number of training pairs, the grain structures and micromagnetically computed anchor points along the demagnetization curve.

For such situations a combination of unsupervised and supervised learning is useful. In a first step, we learn low-dimensional representations of the grain structures. This can be done in an unsupervised fashion. In a second step, we learn the mapping from a low-dimensional latent code that represents a grain structure to the anchor points of the demagnetization curve. Since the feature dimensions of the input vector in latent space are much smaller, we can use smaller models to approximate the demagnetization curve: The complexity of the model operating in the latent space reduces since its input dimension is only the latent space dimension. The smaller model, in turn, can be trained with much less data. This concept of combining unsupervised learning for dimensionality reduction and subsequent supervised learning for make predictions is shown in Figure \ref{fig_super}. Creating the latent space by unsupervised learning can be interpreted as automated feature selection \cite{buduma2017fundamentals}. In addition, the combined approach limits the overall model complexity and hence it acts similar as regularization. Regularization reduces overfitting which is widely encountered when increasing the size of the training set.

\begin{table}[!tb]
	\caption{Layout of the encoder and decoder. The first column is the layer name. The second column and third column are is the layer type and activation function, respectively. We use the nomenclature as specified in Keras \cite{Keras}. The first convolution layer and the last convolution layer use a kernel width $4 \times 4$. For all other convolution layers the kernel width is $2 \times 2$. For all convolution layers we use a stride $2 \times 2$. The total number of trainable parameters of the encoder and decoder are $15\,382\,256$ and $14\,338\,801$, respectively.}
	\label{tab_autoencoder}
	\begin{tabular}{l l c r r l} 
		\hline 
		\textbf{Encoder} & & &  & number of &\\ 
		Layer   & (Type)    & Activation & Output shape & parameters & connected to \\ \hline
		input 1 & (Input)   & -          & $160 \times 160 \times 1$  &      0 &  \\
		conv 1  & (Conv2D)  & elu        & $80 \times 80 \times  16$  &    272 & input 1 \\
		conv 2  & (Conv2D)  & elu        & $40 \times 40 \times  32$  &   $2\,080$ & conv 1 \\
		conv 3  & (Conv2D)  & elu        & $20 \times 20 \times  64$  &   $8\,256$ & conv 2 \\
		conv 4  & (Conv2D)  & elu        & $10 \times 10 \times 128$  &  $32\,896$ & conv 3 \\
		conv 5  & (Conv2D)  & elu        & $ 5 \times  5 \times 256$  & $131\,328$ & conv 4 \\
		flatten & (Flatten) & -          &                     6400   &      0 & conv 5 \\
		dense 1 & (Dense)   & elu        &                     2048   & $13\,109\,248$ & flatten \\
		mean    & (Dense)   & linear     &                      512   &  $1\,049\,088$ & dense 1\\
		variance & (Dense)  & linear     &                      512   &  $1\,049\,088$ & dense 1\\ \hline 
		\textbf{Decoder} & &	&  &  &\\ \hline
		input 2 & (Input)  & -          &                       512   &     0 & \\
		dense 2 & (Dense)  & elu        &                      2048   &  $1\,050\,624$ & input 2 \\
		dense 3 & (Dense)  & elu        &                      6400   & $13\,113\,600$ & dense 2 \\
		reshape & (Reshape) & -         &  $ 5 \times  5 \times 256$  &            0   & dense 3 \\
		convtr 1 & (Conv2DTrans) & elu  &  $10 \times 10 \times 128$  &  $131\,200$    & reshape \\
		convtr 2 & (Conv2DTrans) & elu  &  $20 \times 20 \times  64$  &  $32\,832$     & convtr 1 \\
		convtr 3 & (Conv2DTrans) & elu  & $40 \times 40 \times  32$   & $8\,224$       & convtr 2 \\
		convtr 4 & (Conv2DTrans) & elu  & $80 \times 80 \times  16$   & $2\,064$ & convtr 3 \\
		convtr 5 & (Conv2DTrans) & tanh  & $160 \times 160 \times 1$  &     257 & convtr 4 \\ \hline
	\end{tabular}
\end{table}

We apply a variational autoencoder \cite{kingma2013auto,rezende2014stochastic} for dimensionality reduction. Autoencoders learn to copy their inputs to their outputs. Thereby they learn representing the input state in lower dimensionality. Autoencoders consist of several layers of neurons. The layers are grouped into an encoder and a decoder. The encoder maps the input to two vectors the mean and the variance of the latent code which define a probability distribution over the latent space. The decoder takes a point randomly sampled from this distribution and reconstructs the input. Variational autoencoders create a continuous latent space. Points which are close in latent space will give similar reconstructions \cite{chollet2021deep}.

The encoder and decoder networks are almost symmetric. From the inputs to the
mean and variance vector the number of units decreases from layer to
layer (encoder); from the latent code to the outputs (decoder), the number of
units increases from layer to layer. For encoding the granular structure, the input is the $x$-component of the anisotropy direction on the computational grid. The shape of the input is $160 \times 160 \times 1$. The dimension of the mean and variance vector is 512. The layouts of the encoder and decoder are given in Table \ref{tab_autoencoder}. Convolution layers learn local patterns in a small two-dimensional window whose size is defined by the kernel width. The
distance between two successive windows is called stride. With a $2 \times 2$
stride each convolution layer reduces the number of features by a factor
of 1/2. The activation function determines the output of each unit of a
layer. Clevert and co-workers \cite{clevert2015fast} show that the exponential linear unit
(elu) speeds up learning of autoencoders.  

\begin{table}[!tb]
	\caption{Layout of parameter sharing neural network for the multi-target regression of hysteresis parameters $M_1, M_\mathrm{r}, H_\mathrm{k,90}, H_\mathrm{k,80}, H_\mathrm{k,70},$ and $H_\mathrm{c}$. The first column is the layer name. The second column and third columns are is the layer type and activation function, respectively. We use the nomenclature as specified in Keras \cite{Keras}. Please note that the regression head is repeated for each target value. The total number of trainable parameters is $1\,502\,006$. }
	\label{tab_predictor}
	\begin{tabular}{l l c r r l} 
		\hline 
		\textbf{Shared layers} & & &  & number of &\\ 
		Layer          & (Type)    & Activation & Output shape & parameters & connected to \\ \hline
		input          & (Input)  & -          &  512 &            0 &  \\
		dense shared 1 & (Dense)  & elu        &  512 &   $262\,656$ & input \\
		dense shared 2 & (Dense)  & elu        & 1024 &   $525\,312$ & dense shared 1 \\
		dense shared 3 & (Dense)  & elu        &  512 &   $524\,800$ & dense shared 2 \\
		dense shared 4 & (Dense)  & elu        &  256 &   $131\,328$ & dense shared 3 \\
		dense shared 5 & (Dense)  & elu        &  128 &    $32\,896$ & dense shared 4 \\
		shared output  & (Dense)  & elu        &   64 &     $8\,256$ & dense shared 5 \\ \hline 
		\textbf{Regression head} & &	&  &  &\\ \hline
		dense head 1   & (Dense)  & elu        &   32 &     $2\,080$  & shared output \\
		dense head 2   & (Dense)  & elu        &   16 &        $528$  & dense head 1  \\
		dense head 3   & (Dense)  & elu        &    8 &        $136$  & dense head 2 \\
		dense head 4   & (Dense)  & elu        &    4 &         $36$  & dense head 3 \\
		dense head 5   & (Dense)  & elu        &    2 &         $10$  & dense head 4 \\
		target         & (Dense)  & linear     &    1 &          $3$  & dense head 5 \\ \hline
	\end{tabular}
\end{table}

The free parameters of the variational autoencoders are trained with two loss functions, the reconstruction loss and the regularization loss \cite{chollet2021deep}. We use the mean squared error to bring the reconstructed input close to the original. The Kullback–Leibler divergence between the target 
Gaussian distribution and the actual distribution of the latent code forces the encoder output towards a normal distribution.

\begin{figure}[!ht]
	\centering
	\includegraphics[scale=0.43]{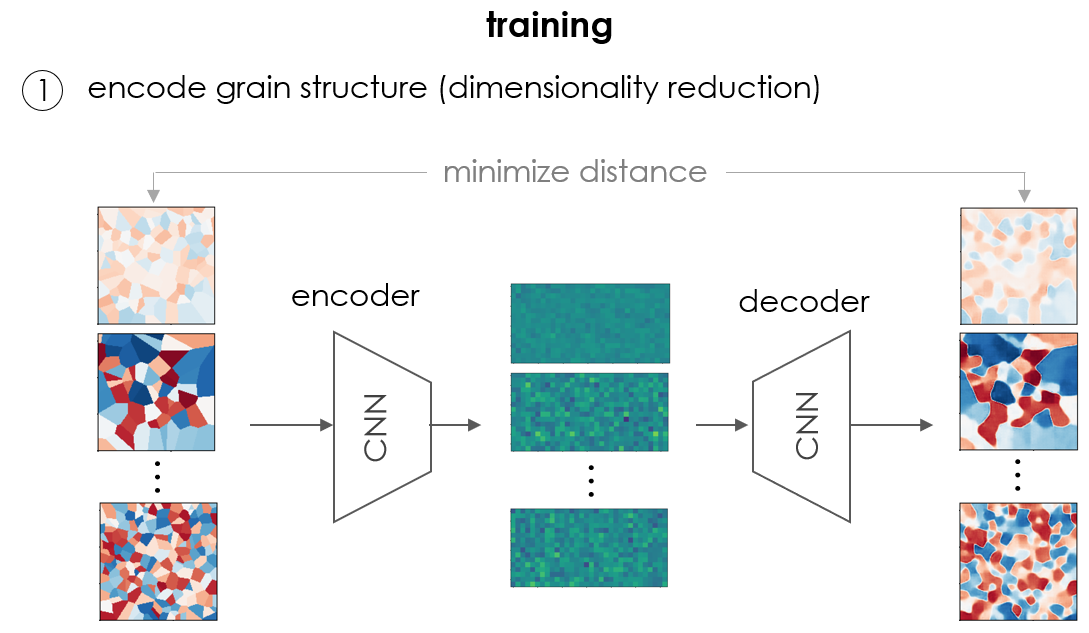}
	\includegraphics[scale=0.45]{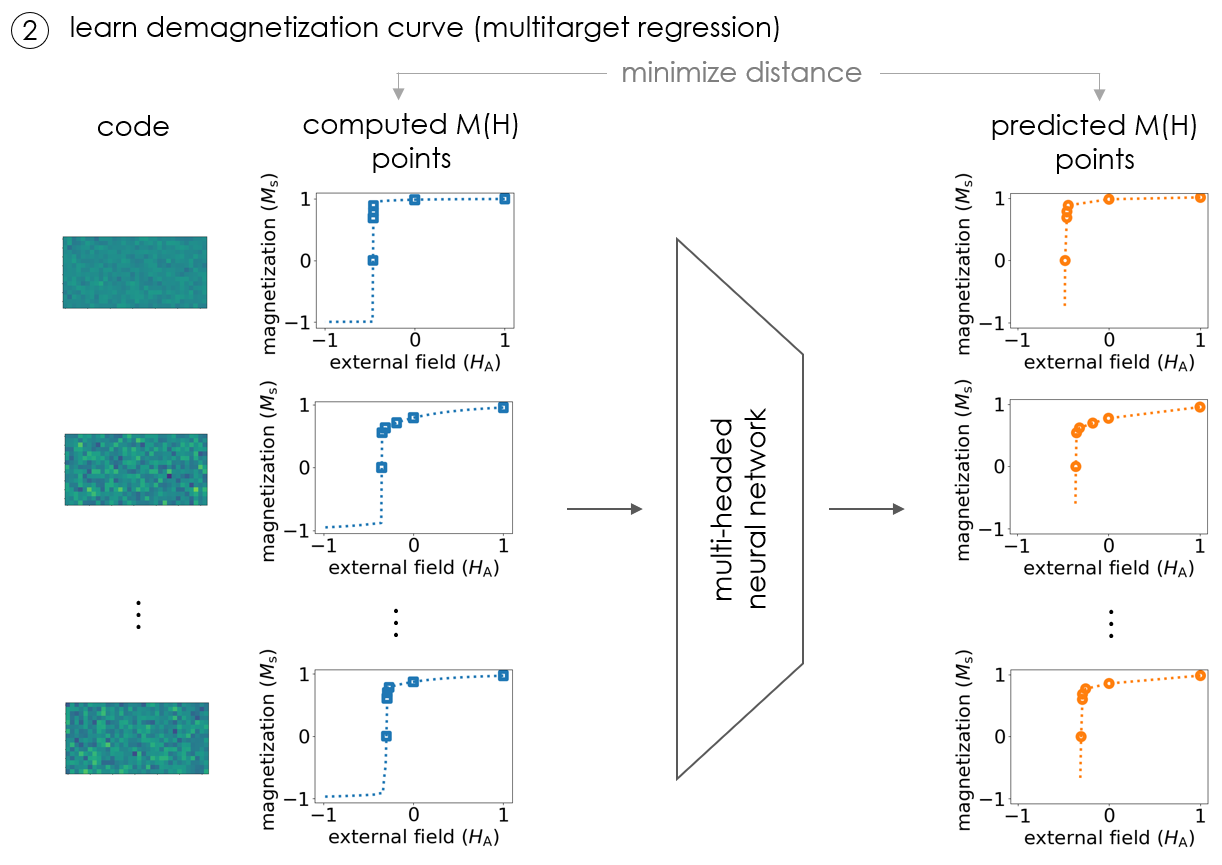}
	\includegraphics[scale=0.5]{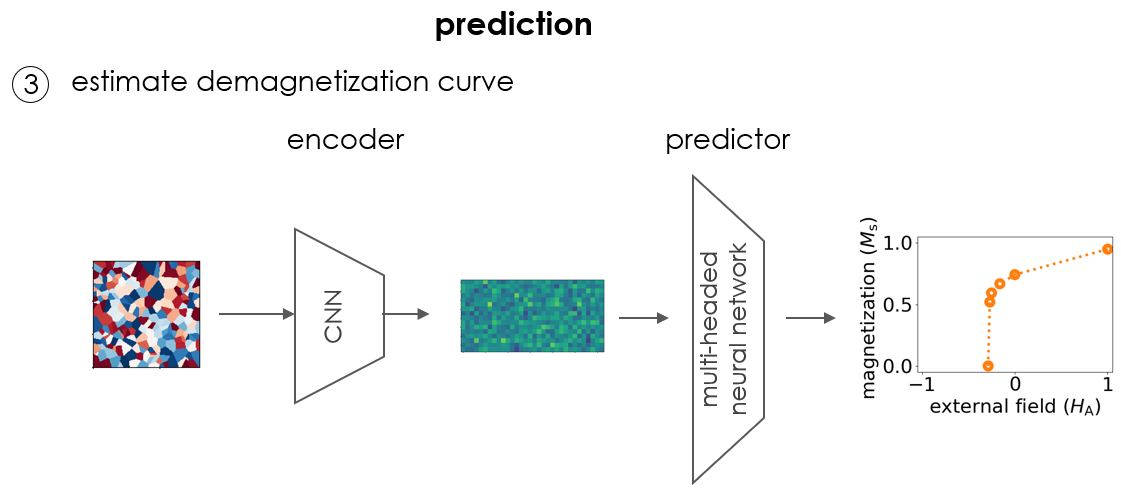}
	\caption{\label{fig_toolchain} The estimation of the demagnetization curve of nanocrystalline magnets from the grain structures requires three steps. In the first step, granular structures are encoded with a low-dimensional latent code. The encoder and decoder are convolutional neural networks (CNN). The second step is the training of multi-target regressor, which receives encoded grain structures as input and gives the anchor points of the demagnetization curve as ouput. In the third step a granular grain structure is first transformed into low-dimensional latent code from which the demagnetihzation curve is predicted. In the images of the grain structure the color maps the $x$-component of the anisotropy direction: -1 red, +1 blue. The latent codes are depicted as image with $32 \times 16$ squares.  } 
\end{figure}

\subsection{Multi-target regression with parameter-sharing deep neural networks}
    
The prediction of the anchor points along the demagnetization curve is a multi-target regression. We want to predict $M_1, M_\mathrm{r}, H_\mathrm{k,90}, H_\mathrm{k,80}, H_\mathrm{k,70},$ and $H_\mathrm{c}$ from the encoded microstructure. In principle, it is possible to train six independent neural networks for the prediction of the six target values. However, Reyes and co-workers \cite{reyes2019performing} showed that parameter-sharing improves the performance of multi-target neural network regression. Such networks contains multiple parts: \emph{A set of layers which is shared} by each target. The shared layers are followed by individual set of layers for each target which we call \emph{regression head}. By sharing network parameters inter-target relationships can be learned. The parameters of the regression head learn to adapt the predictor to the specifics of each target. In most test cases such parameter-shared neural networks out performed independent networks for each target \cite{reyes2019performing}. 

Table \ref{tab_predictor} gives the layout of the multi-headed neural network for prediction of the anchor points. The input is the encoded grain structure. The outputs are $M_1, M_\mathrm{r}, H_\mathrm{k,90}, H_\mathrm{k,80}, H_\mathrm{k,70},$ and $H_\mathrm{c}$. Please note that, through the use of the low-dimensional latent code as input, the total number of trainable parameters is much smaller than that for the autoencoder. Therefore, a small set  input-output pairs might be sufficient to train the predictor.

We apply four-fold cross validation to determine the number of passes through the entire training set (epochs) \cite{chollet2021deep}.

\subsection{Machine learning tool chain}

In the following we briefly summarize the steps required to estimate demagnetization curves from the granular structure of nanocrystalline permanent magnets. We need to train two neural networks: The variational autoencoder for the creation low dimensional latent codes of granular structures and the multi-target regressor for the prediction of the anchor points along the demagnetization  curve. 
\begin{enumerate}
	\item 
	Dimensionality reduction: Granular structures are sampled as discussed in section \ref{sec_grains}. Unsupervised learning trains the weights and biases of the encoder and decoder network. With the encoder a low dimensional latent code can be computed for each granular structure. With the decoder the original grain structure can be reconstructed.
	\item
	Learning the demagnetization curve: Input-output pairs consisting of encoded grain structures and computed anchor points along the demagnetization curve are used to train a multi-headed neural network.
	\item
	Estimation of the demagnetization curve: A grain structure is created, it is encoded into a low-dimensional latent code, and the neural network regressor is used to predict the demagnetization curve, which is approximated with six anchor points.
\end{enumerate}

\begin{figure}[!t]
	\centering
	\includegraphics[scale=0.54]{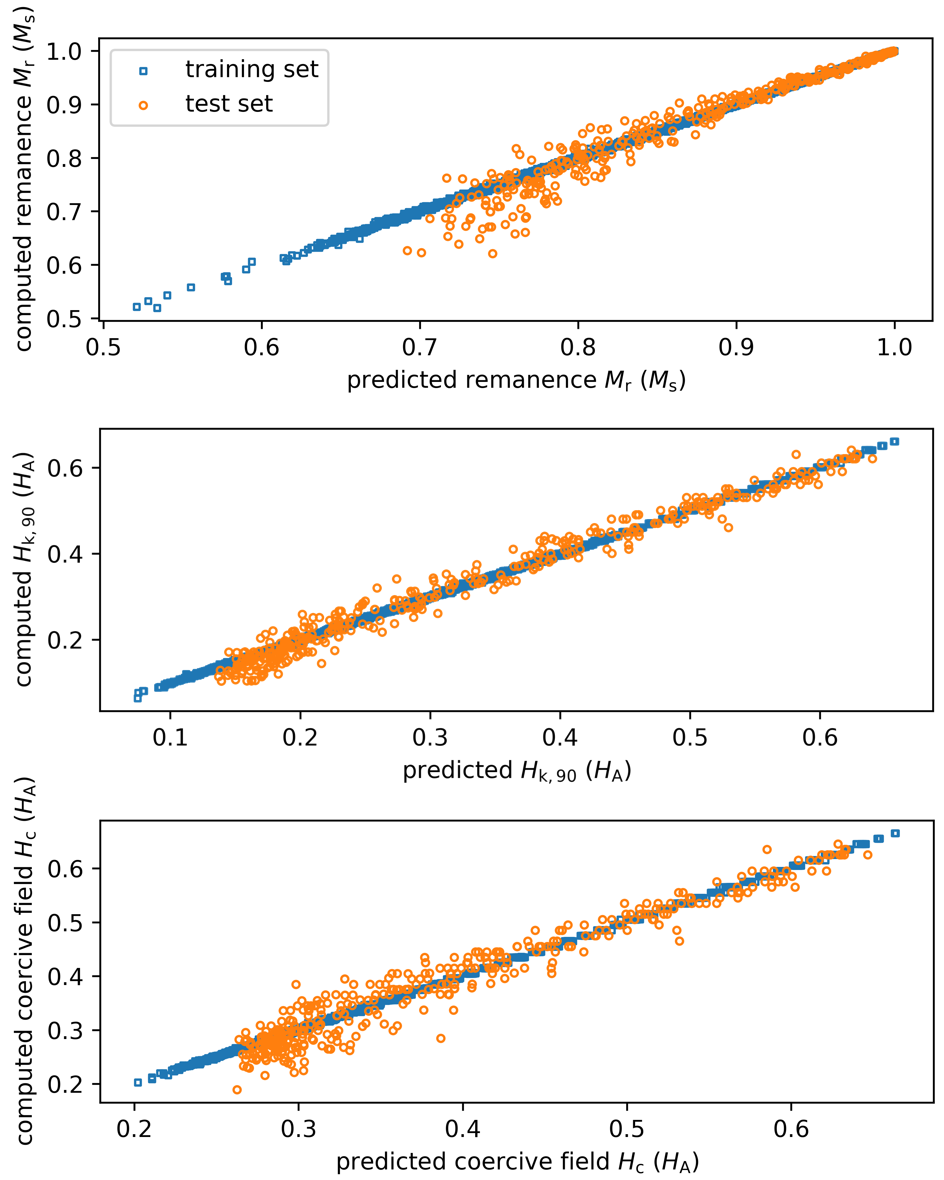}
	\caption{\label{fig_resplots} Computed versus predicted values for key figures of merits of a permanent magnet: The remanence $M_\mathrm{r}$, the knee field $H_\mathrm{k,90}$, and the  coercive field $H_\mathrm{c}$. We can see that the training data are well fitted (blue squares; no underfitting) and the test error is small within the whole range of values (orange circles; no overfitting).} 
\end{figure}

\begin{table}[!b]
	\caption{Mean absolute errors of the predictions for key material properties predicted for the test set. Please see section \ref{sec_grains} for the definition of the quantities. $H_\mathrm{A} = 2 K_1 / (\mu_0 M_\mathrm{s})$ is the anisotropy field.}
	\label{tab_errors}
	\begin{center}
	\begin{tabular}{l c l} 
	\hline 
	property & mean absolute error & in units of \\ \hline
	$M_1$              & 0.0030    & $M_\mathrm{s}$ \\
	$M_\mathrm{r}$     & 0.0152    & $M_\mathrm{s}$ \\
	$H_\mathrm{k,90}$  & 0.0184    & $H_\mathrm{A}$\\
    $H_\mathrm{k,80}$  & 0.0245    & $H_\mathrm{A}$ \\
    $H_\mathrm{k,70}$  & 0.0234    & $H_\mathrm{A}$ \\
    $H_\mathrm{c}$     & 0.0206    & $H_\mathrm{A}$ \\ \hline
\end{tabular}
\end{center}
\end{table}

\begin{figure}[!t]
	\centering
	\includegraphics[scale=0.54]{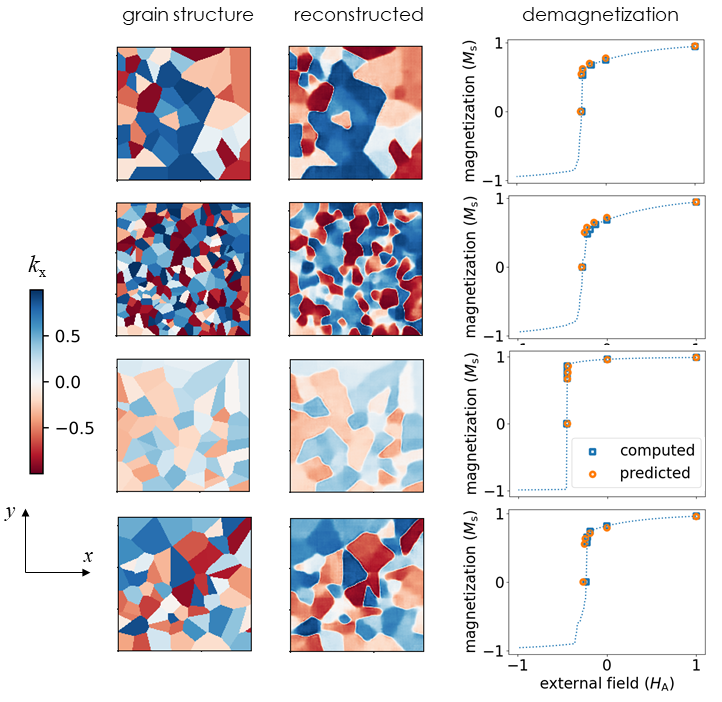}
	\caption{\label{fig_results} Samples from the test set. The columns give the original grain structure, the grain structure decoded from the latent code, and the predicted anchor points along demagnetization curve. In the images of the grain structure the color maps the $x$-component of the anisotropy direction: -1 red, +1 blue. The blue squares and the orange dots refer to the computed and predicted values, respectively. The external field is applied parallel to the $y$-axis. } 
\end{figure}

The machine learning tool chain is shown in Figure \ref{fig_toolchain}.

\section{\label{sec_results}Results}

\subsection{\label{sec_results_hyst}Prediction of hysteresis properties}

We randomly sampled $320\,000$ granular structures using a Voronoi-tesselation as described in section \ref{sec_grains}. The variational autoencoder was trained with 400 passes through the training set (epochs). The batch size was 64. For optimization we applied the Nadam optimizer \cite{dozat2016incorporating} with an initial learning rate of $10^{-4}$. Nadam is a gradient descent optimization algorithm which is supposed to converge quickly. The learning rate determines the step size of the algorithm. 

The orignal grain structure, the latent code (mean output vector of the encoder), and the resonstructed grain structure are shown in Figure \ref{fig_toolchain}.1 for some samples. Pairs of original grain structures and their reconstructions are also given in Figure \ref{fig_results}. The color denotes the $x$-component of the anisotropy direction. The reconstructed granular structures are blurry. This is a well-known drawback of variational autoencoders \cite{goodfellow2016deep}. In our machine learning tool chain we only use the latent code and do not need the reconstructed grain structure except for computing the losses during training of the autoencoder. Indeed, the latent codes, which have been constructed with the variational autoencoder, can be used successfully to predict demagnetization curves as we will show later in this section.

For training and testing the multi-headed neural network we encoded 3977 randomly sampled grain structures using the procedure describe in  \ref{sec_grains}. For these structures we computed the demagnetization curves using the intrinsic material properties of Nd$_2$Fe$_{14}$B ($A = 8 \times 10^{-12}$~A/m \cite{coey_magnetism_2010}, $\mu_0 M_\mathrm{s} = 1.61$~T \cite{kronmuller2003micromagnetism}, and $K_1 = 4.2 \times 10^6$~J/m$ ^3$ \cite{durst1986determination}, the anisotropy field $H_\mathrm{A} = 2 K_1 / (\mu_0 M_\mathrm{s})$ is 6.6~T/$\mu_0$). We computed the equilibrium magnetic states for subsequent decreasing external fields, starting with $H_\mathrm{ext} = H_\mathrm{A}$. The external field was decreased in steps of $0.01 H_\mathrm{A}$ until $H_\mathrm{ext} = - H_\mathrm{A}$ was reached. From the computed demagnetization curve the values $M_1, M_\mathrm{r}, H_\mathrm{k,90}, H_\mathrm{k,80}, H_\mathrm{k,70}$ and $H_\mathrm{c}$ were derived. The latent code of the grain structure and these six values were the input-output pairs for training and testing the multi-headed neural network regressor. 

We used 3607 pairs for training and 370 pairs for testing. For optimization we used the Nadam optimizer with an initial learning rate of $10^{-4}$. We trained the network for 753 epochs with a batch size of 16. The mean absolute errors for $M_1, M_\mathrm{r}, H_\mathrm{k,90}, H_\mathrm{k,80}, H_\mathrm{k,70},$ and $H_\mathrm{c}$ are listed in Table \ref{tab_errors}. Figure \ref{fig_resplots} shows the computed versus the predicted values for the remanent magnetization, the knee field, and the coercive field. 

Finally, we present some samples randomly taken from the test set in Figure \ref{fig_results}. The results show that the remancence, the coercive field, and shape of the demagnetization curve change with the microstructure. The comparision of the computed demagnetization curve with the predicted anchor points shows that machine learning successfully captures the influence of the microstructure. 

\begin{figure}[!t]
	\centering
	\includegraphics[scale=1.1]{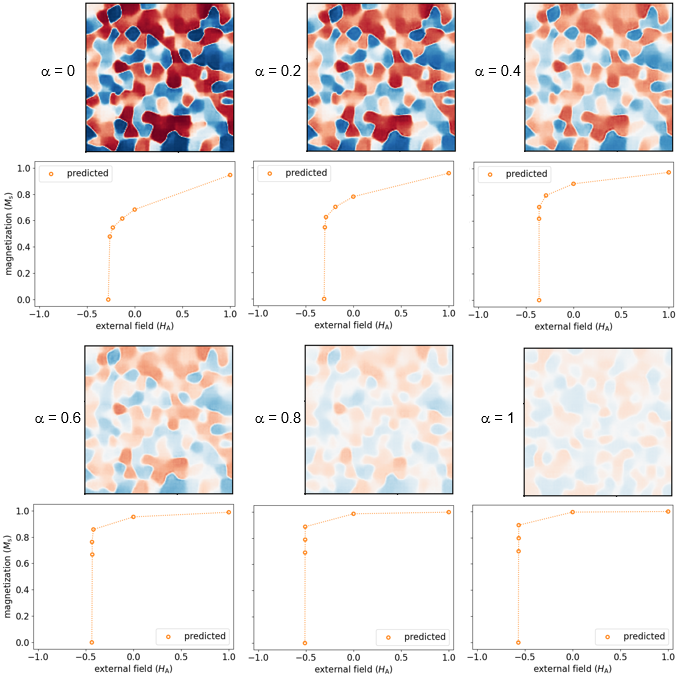}
	\caption{\label{fig_morphing} Morphing between microstructures with different degree of alignement: Linear interpolation between two points in latent space gives the decoded microstructures. The color maps the $x$-component of the anisotropy direction: -1 red, +1 blue.  The external field is applied parallel to the $y$-axis. The demagnetization curves are predicted from the respective latent code with the neural network regressor.} 
\end{figure}

\begin{figure}[!t]
	\centering
	\includegraphics[scale=1.1]{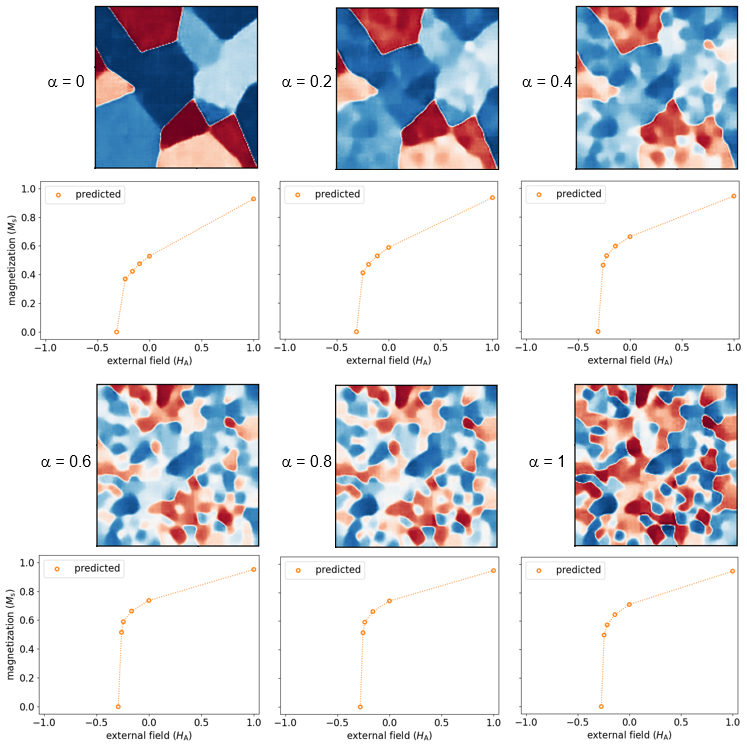}
	\caption{\label{fig_morphing_size} Morphing between microstructures with different grain size distributions: Linear interpolation between two points in latent space gives the decoded microstructures. The color maps the $x$-component of the anisotropy direction: -1 red, +1 blue.  The external field is applied parallel to the $y$-axis. The demagnetization curves are predicted from the respective latent code with the neural network regressor.} 
\end{figure}

\subsection{\label{sec_results_latent}Microstructure morphing}

The latent space of variational autoencoders is highly continuous. To generate a new microstructure and predict its hysteresis properties, we 
\begin{enumerate}
\item perform arithemtic with vectors in the latent space,
\item decode the new latent code to real space, and
\item apply the neural network regressor to estimate the demagnetization curve. 
\end{enumerate} 

Here we give an example for microstructure morphing \cite{foster2019generative}. We pick two points in the latent space, $z_A$ and $z_B$, that represent two microstructure. Then we explore how the hysteresis properties change if we gradually change the microstructure from $A$ to $B$. We walk along the straight line 
\begin{equation}
	\label{eq_latent}
	z_{\alpha} = (1-\alpha) z_A + \alpha z_B
\end{equation}  
between the locations $z_A$ and $z_B$ in latent space. We apply the decoder and the neural network predictor to the latent code $z_\alpha$  for computing the real space image and the demagnetization curve of the intermediate microstructure, respectively. 

In a first example, we picked microstructure $A$ with  $n_g = 197$ and $\theta = 176^\mathrm{o}$ and microstructure $B$ with $n_g = 252$ and $\theta = 18^\mathrm{o}$. Here $n_g$ is the number of grains and $\theta$ is the maximum possible misalignment angle of a grain (see section \ref{sec_grains}). The two structures mainly differ by their degree of alignment, ranging from an isotropic magnet ($A$) to an oriented magnet ($B$). The larger $\alpha$, the more aligned is the magnet.

Figure~\ref{fig_morphing} shows the reconstructed microstructure and the predicted demagnetization curves was we walk along the line between $z_A$ and $z_B$ in latent space. It is clearly visible that an increased alignment enlarges the remanent magnetization as well as the coercivity.  Also the loop squareness increases with increasing alignment.  

In a second example, interpolated between two isotropic magnets with different grain size distributions. We chose microstructure $A$ with  $n_g = 16$ and $\theta = 178^\mathrm{o}$ and microstructure $B$ with $n_g = 256$ and $\theta = 175^\mathrm{o}$. The average grains size of sample $A$ is about 80~nm and that of sample $B$ is about 20~nm. The grain size decreases with increasing $\alpha$. Figure~\ref{fig_morphing_size} shows the reconstructed microstructure and the predicted demagnetization curves from microstructure morphing between isotropic magnets with large and small grains. As expected from previous data in literature \cite{fukunaga1992effect,schrefl1994remanence} the remanence and the loop squareness increase with decreasing grain sizes. The improvement in squareness of the demagnetization curve occurs for $0 \le \alpha \le 0.6$. 

The above examples show the power of generative models. New microstructures can be generated and their properties can be predicted. Figure~\ref{fig_morphing_size} also shows how a simple linear interpolation in latent space can lead to new structures that are unexpected. For $\alpha = 0.2$ and $\alpha = 0.4$, the generated structures show large grains which break up into smaller subgrains. It is well known, that simple linear interpolation between two points in latent space may give locations with low probability in the distribution of the model's training data \cite{white2016sampling}. However, for applications such as the search for new structures with superior properties, it may be advantageous to generate new samples which clearly differ from the discrete set of training samples. 

\section{\label{sec_outlook}Summary and outlook}

This work is a proof of concept that it is possible to estimate the demagnetization curve of permanent magnets from the microstructure with data-driven machine learning. We used synthetically generated grain structure and micromagnetically computed demagnetization curves. The granular structure of the magnet was successfully encoded by with a low-dimensional latent space which was constructed by a variational autoencoder. Training of the autoencoder was unsupervised. Thus, there was no resource related restriction on the size of the training set since synthetic structures could be generated fast. In contrast computing the demagnetization curve is time consuming. Because the latent code captured key features of the microstructure, the number of trainable parameters of a multi-target neural network regressor for the demagnetization properties could be kept small. The small network size reduced the required amount of labeled training data.  

Though this was a purely computational work, we can imagine that microstructural images such as electron backscatter diffraction (EBSD) data and experimentally measured magnetic properties may be used for training the model. Furthermore, data fusion that merges experimental and computed training data is possible. Alternatively to synthetically generated microstructure, microstructure generation may be based on EBSD data  \cite{gusenbauer2019automated,gusenbauer2020extracting}.

A variational autoencoder is a generative model. We demonstrated how vector operations in latent space can be used to create new  microstructures. The methodology introduced in this paper can be applied for inverse material design. To find magnets that have demagnetization curves of with given properties, one can apply a genetic algorithm using the neural network predictor as the forward model for the evaluation of the objective functions. 

\section*{Acknowledgment}
The financial support by the Austrian Federal Ministry for Digital and Economic  Affairs, the National Foundation for Research, Technology and Development and the Christian Doppler Research Association is gratefully acknowledged. 
L.E. acknowledges support by the Austrian Science Foundation (FWF) under grant No. P31140-N32.

\bibliographystyle{elsarticle-num-names}
\bibliography{references.bib}

\end{document}